\documentclass[10pt]{article}

\usepackage[a4paper, margin=0.75in]{geometry}
\usepackage{authblk}
\usepackage{graphicx}
\usepackage{epstopdf, epsfig}	
\usepackage{amsmath}
\usepackage{amssymb}
\usepackage[round,sort&compress,comma,authoryear]{natbib}
\usepackage{float}
\usepackage{hyperref}
\usepackage{lscape}
\usepackage{newtxtext}
\usepackage{newtxmath}
\usepackage{natbib}
\usepackage{hyperref}
\hypersetup{
    colorlinks = true,
    urlcolor   = blue,
    citecolor  = black,
}

\newcommand{\rectangle}{\fboxsep0pt\fbox{\rule{0.6em}{0pt}\rule{0pt}{0.9ex}}}

\title{{Liquid inertia versus bubble cloud buoyancy in {circular} plunging jet experiments}} 

\author[1]{Narendra Dev}
\author[1]{{J. John Soundar Jerome}}
\author[1]{Hélène Scolan} 
\author[1, $\dagger$]{Jean-Philippe Matas}
\affil[1]{Univ Lyon, Univ Claude Bernard Lyon 1, CNRS, Ecole Centrale de Lyon, INSA Lyon, LMFA, UMR5509, 69622 Villeurbanne France}
\affil[$\dagger$]{\href{emailto:jean-philippe.matas@univ-lyon1.fr}{jean-philippe.matas@univ-lyon1.fr}}

\date{}
\begin{document}
\maketitle

\begin{abstract}
When a liquid jet plunges into a pool, it can generate a bubble-laden jet flow underneath the surface. This common and simple {phenomenon}  is investigated experimentally {for circular jets} to illustrate and quantify the role played by the net gas/liquid void fraction on the maximum bubble penetration depth. It is first shown that an increase in either the impact diameter or the jet fall height to diameter ratio at constant impact momentum leads to a reduction in the bubble cloud size. By systematically measuring the local void fraction using optical probes in the biphasic jet, it is then demonstrated that this effect is a direct consequence of the increase in the air content within the cloud. A simple momentum balance model, including only  {inertia and} the buoyancy force, is shown to predict the bubble cloud depth without any fitting parameters. Finally, a Froude number based on the bubble terminal velocity, the cloud depth, and also the net void fraction is introduced to propose a simple criterion for the threshold between the inertia-dominated and buoyancy-dominated regimes. 
\end{abstract}

\section{Introduction}
\label{sec:intro}
The impact of a plunging jet on the free surface of a pool of the same or different liquid above a critical velocity transports ambient gas into the continuous liquid phase forming a cluster of bubbles, a bubble cloud {\citep{lin1966gas, bonetto1994analysis, zhu2000mechanism, lorenceau2004air}}. This phenomenon is widely encountered in industrial applications like the stirring of chemicals \citep{mckeogh1981air} or hydroelectric applications {\citep{guyot2019contribution}}, and in nature, as in breaking waves and cascades \citep{chanson2002hydraulics, kiger2012air}.

\begin{figure}
	\begin{center}
		\epsfig{file=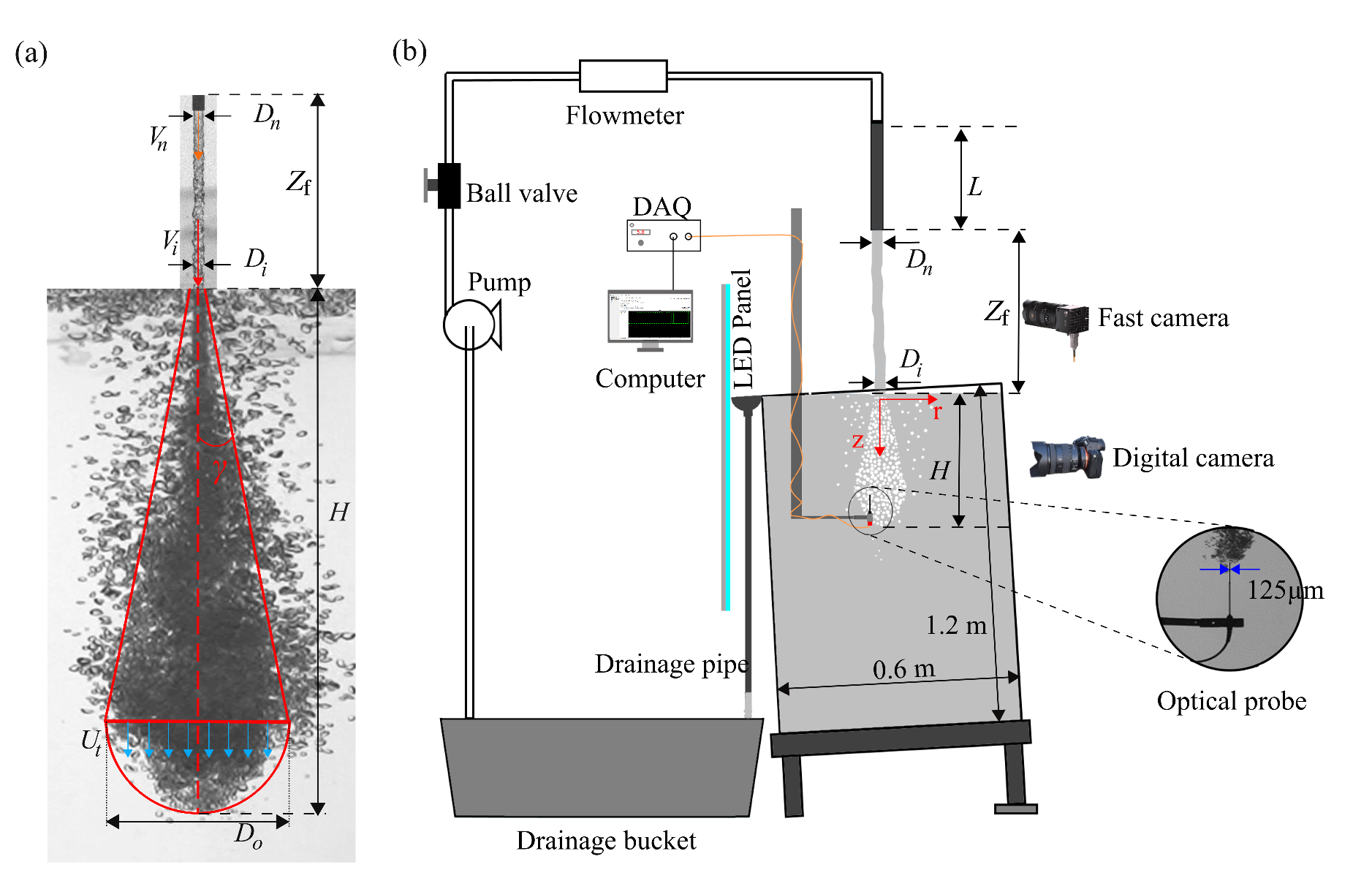, width=1\textwidth,keepaspectratio=true}
	\end{center}
    \caption{(a) Schematic of control volume of the bubble cloud, and (b)  
    Schematic of the experimental setup illustrating flow lines to generate bubble cloud, back-lighting imaging setup and void fraction measurement setup with optical probes.}
\label{fig:schematic}
\end{figure}

{Previous studies discussed the inception of air bubbles below jet impact, and proposed models for the air entrainment mechanism based on the morphology in {circular and \citep{mckeogh1981air,sene1988air,el2002measurements}, planar jets \citep{cummings1999experimental,bertola2018physical} , and multi-droplet streams \citep{speirs2018water}.} \citet{kiger2012air} reviewed the air-entrainment mechanism for laminar, turbulent and disintegrated jets of different viscosity}. { Once the bubble entrainment conditions are reached,} the maximum penetration depth of the bubble cloud, denoted as $H$ in figure \ref{fig:schematic}(a), is an essential parameter to model in various applications \citep{clanet1997depth}. Indeed a change in $H$ will affect the volume of the biphasic region and alter the gas mixing rate in chemical industries. In hydroelectric power plants, $H$ is critical in designing dams to prevent erosion and structural weakening caused by the incoming biphasic jet hitting the riverbed. {As can be inferred from Table \ref{tab:PastJetExp},} previous studies have addressed the bubble cloud formation over a wide range of scales \citep{ bin1993gas, clanet1997depth, chirichella2002incipient, roy2013visualisation, qu2013experimental, harby2014experimental,kramer2016penetration, miwa2018experimental, guyot2020penetration}. {Evidence for strong scale effects exist not only regarding bubble cloud size but also for bubble count rate or void fraction \citep{chanson2004physical}.}  Whereas numerous empirical correlations for the cloud depth can be found in the literature depending on their size from a few centimetres to about a few metres, the underlying physics has only been recently elaborated.

\begin{table}
\centering
\begin{center}
  \begin{tabular}{cccccccc}
         \hline
         &  & & &	&\\
      Author(s) (year)	&Jet and Nozzle &$D_n$	&$V_n$	&$Z_f$	&Re $= {\rho V_n D_n}/{\mu}$    \\[2pt]
         &geometry  &(mm) &(m s$^{-1}$)	&(cm)	&(max)\\[2pt]
         \hline
         &  & & &	&\\
\citet{lin1966gas}  & $\downarrow$ $\medcirc$  & 2-8  &0.8-2  & 1-17 &  $2.5 \times 10^3 $  \\

\citet{van1973surface, van1976jet}  & $\swarrow$ $\medcirc$ & 2.85-10  & 3-25 & 10-40  &  $1.12 \times 10^5 $  \\

\citet{van1981water}   & $\downarrow$  $\medcirc$  & 4-30  & 4-10  & 20-85  &$2.8\times 10^5 $    \\

\citet{mckeogh1981air}  & $\swarrow$ $\downarrow$ $\medcirc$  & 6-30  & 1-7   & 50-500  &  $1.1 \times 10^5 $   \\
 
\citet{clanet1997depth} & $\downarrow$ $\medcirc$  & 0.24-2.16  & 1.9-18  & 0.7-4  &   $1.4 \times 10^4$  \\

\citet{cummings1999experimental} & $\downarrow$ $\rectangle$   & $3-10$  & 1.1-2   & 0.5-18   &    $1.16 \times 10^4$   \\

\citet{chanson2003air} & $\downarrow$ $\medcirc$    & 14-25  & 0.5-5   & 5-20   &    $1.6 \times 10^5$           \\

\citet{chanson2004physical} & $\downarrow$ $\medcirc$  & 6.8-25 & 1.7-4.4 & 3-10  &   $1.23 \times 10^5$\\

\citet{harby2014experimental} & $\downarrow$ $\medcirc$   & 4-14 & 2.5-13 & 2.5-30  &   $9 \times 10^4$  
\\
\citet{kramer2016penetration} & $\downarrow$ $\medcirc$  & 13-81.9 &5-7 & 20-100  &   $6.4\times 10^5$ \\
\citet{bertola2018physical} & $\downarrow$ $\rectangle$   & $10.4-11.5$  & 2.4-7.4   &10.4-12.7   &    $9.4 \times 10^4$   \\

\citet{guyot2019depth}  & $\rightleftarrows$ $\medcirc$ & 0.3-2.4  & 1.89-20 & 0.6-4.8 &        $2.4 \times 10^4$\\

\citet{guyot2020penetration}  & $\downarrow$ $\medcirc$   & 0.3-213  & 3-28  & 20-950  &       $3 \times 10^6$                          \\

Present study  & $\downarrow$ $\medcirc$   & 2.7-10  & 2.5-10  & 5-80  &   $1 \times 10^5$  \\          
  \end{tabular}
\end{center}
  
  \caption{Flow conditions in previous works on vertical ($\downarrow$), inclined ($\swarrow$), and oscillating ($\rightleftarrows$) plunging jets from circular ($\medcirc$) and planar ($\rectangle$) nozzles. All authors used water. Data in \cite{lin1966gas} refers also to oil and glycol jets.}
  \label{tab:PastJetExp}
\end{table}
{In this context, \citet{clanet1997depth} developed a jet momentum conservation argument, similar to \citet{suciu1976gas}, but by postulating that (i) the biphasic jet expands with a constant cone half-angle $\gamma=12.5^{\circ}$, as it is well known in turbulent jets \citep{horn1956angle, ervine1987behaviour, l2008random}, and (ii)} {bubbles would escape the biphasic zone as soon as the local advection speed, due to liquid jet below the free-surface,} decreases down to the terminal velocity {$U_t$.} {They thereby derive} a simple prediction for the bubble cloud depth $H_i$:
\begin{equation}
	H_i = \frac{1}{2\tan\gamma} \left( \dfrac{V_{i}D_{i}}{U_{t}} \right),
\label{eq:clanet}
\end{equation}
where $V_i$ is the impact velocity and $D_i$ the impact diameter. Note that there is no adjustable parameter nor any liquid or gas physical properties in equation (\ref{eq:clanet}), except for $U_t$ which is taken equal to $22$~cm/s for all air bubbles in water larger than $1$ mm in diameter \citep{maxworthy1996experiments}. Eq. (\ref{eq:clanet}) shows that in this case, the cloud depth is directly proportional to $V_{i}D_{i}$, which, up to a factor of liquid density $\rho$, is the square root of the impact momentum. \cite{clanet1997depth} obtain a very good agreement for their experiments with micro-jets, of at most {$2.16$~mm} in nozzle diameter. {However, unlike single-phase jets in a neutrally-buoyant environment,} momentum conservation may not be valid for voluminous biphasic jets, for which buoyancy cannot be neglected. {For example, in a $15$~cm deep conical bubble cloud which contains only a small void fraction of air, say $\phi = 10$\%, the outgoing momentum $\dot{Q}_o \sim (1 - \phi) \rho {U_t}^2 \pi H^2\tan^2\gamma$ in Fig. \ref{fig:schematic}(a) is already comparable with the net buoyancy force $F_b \sim \phi \rho g {\pi}H^3\tan^2\gamma/3$ acting on it. Indeed, the largest cloud in experiments by \citet{clanet1997depth} is about $16$~cm deep, and equation (\ref{eq:clanet}) over-predicts the cloud depth data in the previous literature \citep[see Fig. 29 and references within]{bin1993gas} for much deeper clouds}.

{More recently, \citet{guyot2020penetration} extended the above momentum balance model by accounting for buoyancy effects under the assumption that the void fraction within the bubble cloud was uniform.} 
They applied the momentum balance across the truncated conical control volume in Fig. \ref{fig:schematic}(a) to obtain the following equation for the modelled cloud depth $H_b$:
\begin{eqnarray}
	\label{eq:guyot}
 	\underbrace{\rho V_i V_n \dfrac{{\pi} {D_n}^2}{4}}_\text{Incoming momentum flux} &= &\underbrace{\left(1-\bar{\phi}\right) \rho {U_t}^2 {\pi} \left(H_b \tan\gamma + \dfrac{D_i}{2} \right)^2}_\text{Outgoing momentum flux}\\ \notag
&{~} &+\underbrace{\bar{\phi} \rho g \pi \left(\dfrac{1}{3}H_b^3 \tan^2 \gamma + \dfrac{D_i }{2}H_b^2 \tan \gamma + \dfrac{D_i^2}{4} H_b \right)}_\text{Buoyancy force},
\end{eqnarray}
where $D_n$ is the nozzle diameter, $V_n$ the velocity at the nozzle, and $\bar{\phi}$ the constant void fraction.
{Note that the impact} void fraction and diameter have been removed from the impact momentum term on the left-hand side using mass conservation, {$V_nD_n^2 = (1-\Bar{\phi})V_iD_i^2$}.  
For the limiting case of micro-jets {and bubble cloud depth of $10$~cm or lesser}, the buoyancy force might become less important compared to the outgoing momentum flux in Eq. (\ref{eq:guyot}), which then reduces to Eq. (\ref{eq:clanet}). Conversely, in the limit of very large jets{, or massive bubble clouds $H > 20$~cm}, the contribution of the outgoing momentum flux becomes negligible, and Eq. (\ref{eq:guyot}) reduces to the impact momentum being consumed entirely by buoyancy. The fact that a contribution proportional to a surface{-term} is balanced by a contribution proportional to a volume{-term then} leads to $H$ being proportional to $(V_i V_n)^{1/3} D_n^{2/3}${. This} is close to the empirical correlation $H \propto (V_{n}D_{n})^{0.66} $ proposed by \cite{van1981water} and $H \propto  (V_{n}D_{n})^{0.7}$ proposed by \cite{mckeogh1981air}.
{Hence, based on physical arguments, Eq. (\ref{eq:guyot}) elucidates the two most common empirical relations, namely, $H \propto V_n D_n$ and $H \propto \left( V_n D_n \right)^{0.68}$, over a very wide range of jet diameters and all available data in the current literature \citep{bin1993gas, guyot2020penetration}.}

Furthermore, all parameters in the cubic polynomial for $H_b$ are known a priori, except for the average gas/liquid void fraction $\bar{\phi}$. So, \cite{guyot2020penetration} admitted $\bar{\phi} = 15$~\% based on results from past studies \citep{mckeogh1981air,van1981water}, and obtained a relatively good agreement with large-scale experiments. However, they observed a significant dispersion {for the cloud depth both in their data and previous investigations when the jet diameter is large. \cite{guyot2020penetration} attributed} this dispersion to the variations in the void fraction $\bar{\phi}$ {between various experiments since it} is expected to depend strongly on both the jet geometry and dynamics \citep{mckeogh1981air, sene1988air, bonetto1994analysis, zhu2000mechanism}. Therefore, even if the scaling law for the bubble cloud depth is understood to some extent, measurements of the actual values of the void fraction $\phi$ are needed to clarify the role of this quantity on the observed dispersion in the cloud depth data.  The conditions for the transition from jet momentum-dominated to buoyancy-dominated clouds remain to be established as well. In addition, such investigations can provide insights on how, when and where additional features of the bubble cloud such as turbulent dissipation, bubble size distribution, and bubble cloud shape are necessary for dimensioning industrial applications.

{\cite{bin1993gas} lists correlations based on both \textit{global} gas/liquid entrainment ratio measurements, using { gas-hold up techniques}, and \textit{local} gas/liquid void fraction $\phi$ measurements, often using resistivity probes within a bubble cloud. More recent developments on various correlations, along with inception conditions and some mechanistic viewpoints on air-entrainment rate can be found in \citet{kiger2012air}.} 
Past studies evidence that in the developing region, $\phi$ radial profiles consist of two maxima located at the radius of the free jet and a minima at the jet axis {\citep{van1981water, bonetto1993ijmf, cummings1997air, brattberg1998air, chanson2003air, ma2010quantitative}.} \cite{mckeogh1981air} and \cite{van1981water} measured $\phi$ in the fully-developed region and reported that {the peak void fraction $\phi_0$ occurs along the cloud axis, remains almost} constant with depth and then sharply decreases. {However, the current literature lacks measurements of cloud depth $H$ along with measurements of local void fraction $\phi$ in the developed zone of the plume. Such data over various impact momentum and jet fall-height is necessary to help better understand the dynamics of bubble clouds in the buoyancy-dominated regime.}

To address this gap, the objective of the present work is to report {concurrent} {\textit{local}} void fraction $\phi$ and bubble cloud depth $H$ measurements, for {three} {typical} nozzle diameters, and at several impact momentum and {fall-heights} $Z_f$. {Thereby, our aim is two-fold: we examine not only the robustness and the relevance of the momentum balance models \citep{suciu1976gas, clanet1997depth, guyot2020penetration} by incorporating \textit{in-situ} void fraction data into Eqn. (\ref{eq:guyot}), but also its role on the maximum penetration depth of a two-phase plume, by independently varying three major parameters, namely, nozzle diameter, impact momentum, and fall height. Using such a case study, we also try to} propose a simple criterion to determine when the bubble cloud size is controlled by inertia, and when it is controlled by buoyancy.

\section{Methodology}
\label{sec:methodology}
Lab-scale experiments are carried out using a $1.2$~m $\times$ $0.6$~m $\times$ $0.6$~m glass tank which is slightly tilted on one side so that overflowing water is drained through a drainage pipe to the reservoir bucket to keep the water level constant as shown in Fig. \ref{fig:schematic}(b). Pumps supply water from the drainage bucket to the straight injector, from which it is issued as a plunging jet. {Three nozzle diameters $D_n$ are used: $D_n =$  2.7, 8 and 10 mm}. For nozzle diameter $D_n =$  2.7 mm, a series of two centrifugal pumps from  Pan World (NH-100PX and NH-200PS) is used. A heavy-duty centrifugal pump (2KVC AD 45/80M) from DAB Pumps generates the jet from the larger 8 mm {and 10 mm} injectors. The injector length-to-diameter ratio $(L/D_n)$ is kept equal to 50, and the height of fall-to-diameter ratio $Z_f$/$D_n$ is varied from 20 up to 100. Nozzle velocity $V_n$ varies between 1.5 to 12 m/s and is controlled using a flowmeter and ball valve. Impact velocity $V_i$ is deduced using equation $V_i= \sqrt{V_n^2 + 2gZ_f}$. {The present work only considers the cases where the jet is not broken into droplets before impact.}


Backlight imaging is used to film the jet just before the impact bis{illuminated by LED panel using a CMOS fast camera from Ximea (CB262MG) at 300 FPS and 20 $\mu$s shutter speed at 2496 × 2418 resolution.} Images of the biphasic bubble cloud are filmed using a digital camera (Sony a7 III) and a zoom lens (Sony FE 24-70mm F/4 ZA OSS Carl Zeiss, focal length kept at 35 mm) at 50 FPS and bis{200 $\mu$s exposure} at a resolution of 1920 × 1080 pixels. { Using a MATLAB-integrated calibration application, 
lens distortion was found negligible and a constant spatial resolution of 54 and 333 $\mu$m/pixel was used for jet and bubble cloud images, respectively.} The cloud images are analysed {using the open-source freeware \textit{ImageJ} \citep{Schindelin_NatureMethods2012fiji} and algorithms therein for brightness thresholding \citep{Kapur_CompVision1985thresholding,Tsai_1985CompVisionthresholding}, along with an \textit{in-house} MATLAB code to detect and} trace the outer boundaries of the cloud in order to measure $H$. {The mean penetration depth, $H$, is  measured by averaging  over 3000 images. A low frame rate was chosen in order to ensure that the images of the bubble cloud are decorrelated and so that the ensemble average value of the cloud depth $H$ converges faster for a given number of images.} 

{The contours of jet are extracted from images using machine-learning-based object detection models, namely, Grounding DINO \citep{liu2023grounding} and Segment Anything Model (SAM) \citep{kirillov2023segment}. The detected jet edges are then analysed using an in-house MATLAB code to obtain the impact diameter ($D_i$) and the roughness ($\epsilon$). The latter is defined as the average of the RMS values of the lateral departure from the mean edge position, for each side, and is obtained by averaging over $200$ images.}
 
 { The local void fraction $\phi$ is measured using an optical fiber probe (A2 Photonics Sensors) of diameter 125 $\mu m$ mounted  upright on a slender plate (thickness = 4 mm), as shown in the magnified view in Fig. \ref{fig:schematic}(b). This horizontal plate is fixed to a vertical cylindrical rod (diameter = 1.5 cm). 
The probe tip is located at 4.5 cm upstream of this arm. This ensures that obstruction to the incoming flow is minimized as much as possible at the probe tip, where the measurement is carried out. 
The probe assembly is translated in the vertical plane ($r$-$z$ plane) using two motorized linear stages (Igus) having an accuracy of $0.01$ mm. The probe signals are sampled at $250$ kHz for $60$ seconds, a duration which is enough to ensure convergence of the void fraction for all conditions.  The signals are analysed using A2 Photonics software (SO6 v4.7). 
The probe is centered at the jet axis and the measurements are carried out in the radial direction of the cloud with a step size of $2$ mm to obtain radial void fraction profiles. This was repeated at various depths such that $z/H$ = 0.4, 0.6, and $0.8$. Note that when the bubble cloud size is measured, the probe assembly is pulled out of the tank with the help of the translation stage.}

\section{Results and discussion}
\label{sec:results}
 \subsection{Measurement of bubble cloud depth $H$} 

\begin{figure}  

	\begin{center}		\epsfig{file=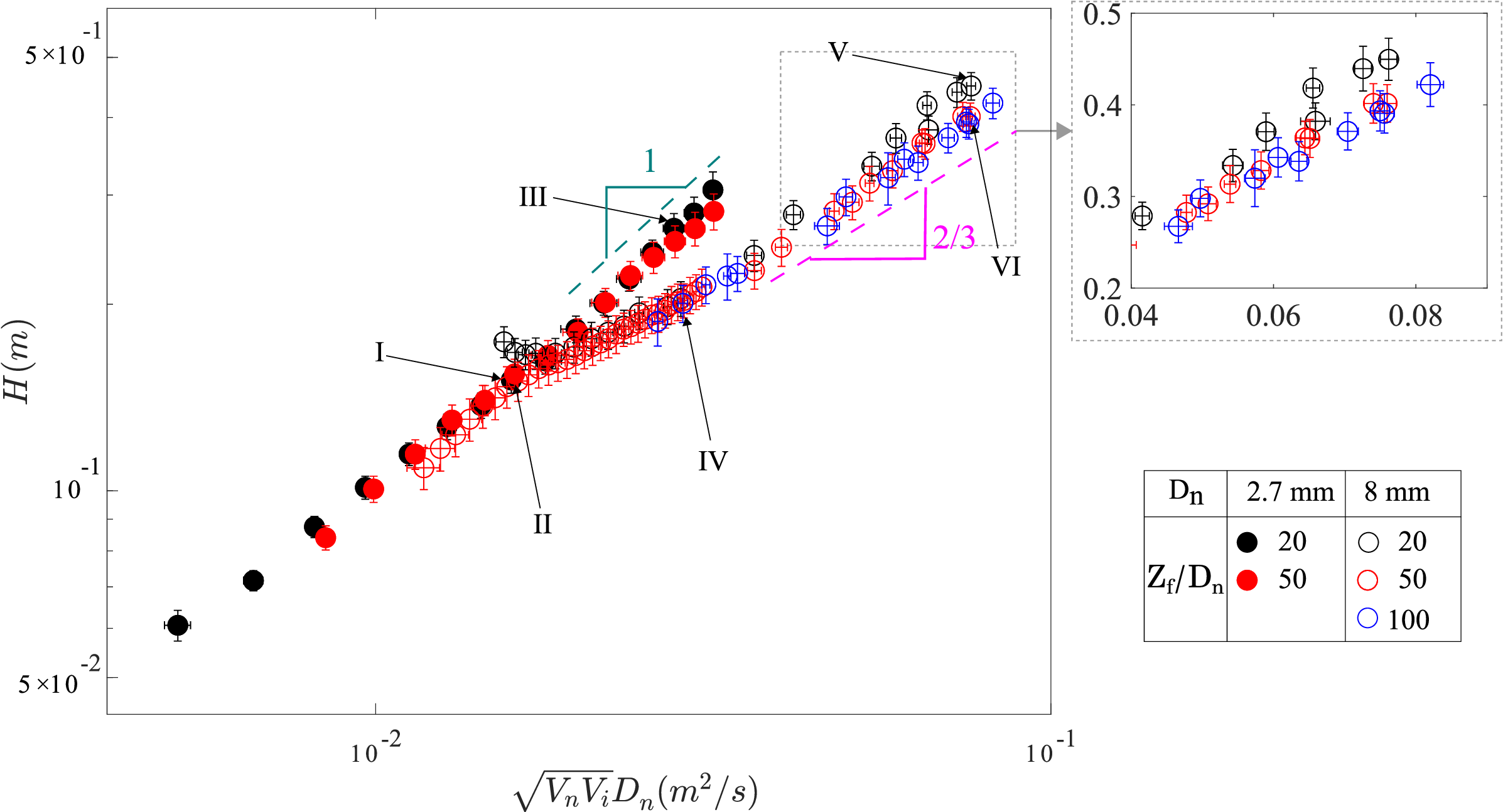,width=1\textwidth,keepaspectratio=true}
	\end{center}
  \caption{Variation of $H$ with square root of impact momentum $\sqrt{V_iV_n}D_n$ for $D_n$ = 2.7 and 8 mm at various $Z_f/D_n$. The solid line shows the trend expected from Eq. (\ref{eq:clanet}). The dashed line shows the 2/3 scaling law expected at large scales. The inset graph on the top-right is a blow-up of the $D_n=8$~mm data showing that increasing $Z_f$ at constant impact momentum leads to a decrease in $H$. {I - VI are the data points corresponding to the flow conditions that are compared in upcoming section.}}
\label{fig:HVsSqrtViVnDn}
\end{figure}
Figure \ref{fig:HVsSqrtViVnDn} illustrates the variation of $H$ for the $D_n=2.7$~mm and $D_n=8$~mm nozzles, as a function of $\sqrt{V_iV_n}D_n$, which {is proportional to the square root of the impact momentum flux, up to a factor $\rho$,} as in equation (\ref{eq:guyot}). {Note that the largest bubble cloud obtained for our conditions is $45$ cm deep, which is significantly smaller than the tank depth ($1.25$~m). We therefore assume that there is no significant pressure gradient caused by confinement effects in our experiments.}  The  bubble cloud size $H$ for jets issued from the $D_n =2.7$~mm nozzle {(red and black discs)} varies linearly with $\sqrt{V_iV_n}D_n$. This is consistent with the model proposed by \cite{clanet1997depth} (solid line), equation (\ref{eq:clanet}), since for this small-scale jet and for the experimental range considered here $V_i\approx V_n$ and $D_i\approx D_n$. When $Z_f/D_n$ is increased for this $D_n =2.7$~mm nozzle, a negligible change in $H$ is observed. For the larger $D_n = 8$~mm nozzle (open symbols), Eq. (\ref{eq:clanet}) accurately predicts the behaviour of $H$ up to $\sqrt{V_iV_n}D_n \approx 0.02$~m$^2$/s.  Beyond this threshold, a transition occurs, and $H$ scales as $(V_iV_n)^{1/3}D_n^{2/3}$, in accordance with the model of \cite{guyot2020penetration} (2/3 power law shown by dashed line). This suggests that beyond a threshold, the buoyancy force becomes dominant over the outgoing momentum flux, leading to a decrease in {cloud depth} $H$ compared to what equation (\ref{eq:clanet}) would predict. Also, Fig. \ref{fig:HVsSqrtViVnDn} {shows} the coexistence of two regimes in {a narrow} range $\sqrt{V_iV_n}D_n= 0.02$ -- $0.03$~m$^2/$s{, wherein the inertia-dominated regime occurs for $D_n = 2.7$~mm while buoyancy forces become important for the bigger nozzle diameter. This is perhaps a signature of the dependence of the transition threshold on the {void fraction} at different jet diameters. Finally, in the buoyancy-dominated zone for $D_n = 8$~mm,} some dispersion in the values of $H$ is observed when $Z_f/D_n$ is varied. As mentioned in the introduction, this dispersion {arises from} the variations in the {net air/water} void fraction, as will be discussed in the next subsection.

\subsection{Local void fraction measurements}
\label{subsec:phi}

The aim of this subsection is to present measurements of the local void fraction $\phi$ within the bubble cloud and discuss them in relation to the bubble cloud size. We first discuss $\phi$ measurements along the jet axis, and then radial profiles of $\phi$ at a given depth within the cloud. Finally, general remarks on the void fraction profiles are presented.


 \subsubsection{Axial measurements}
 \label{subsubsec:phi_axial}

\begin{figure}
	\begin{center}
		\epsfig{file=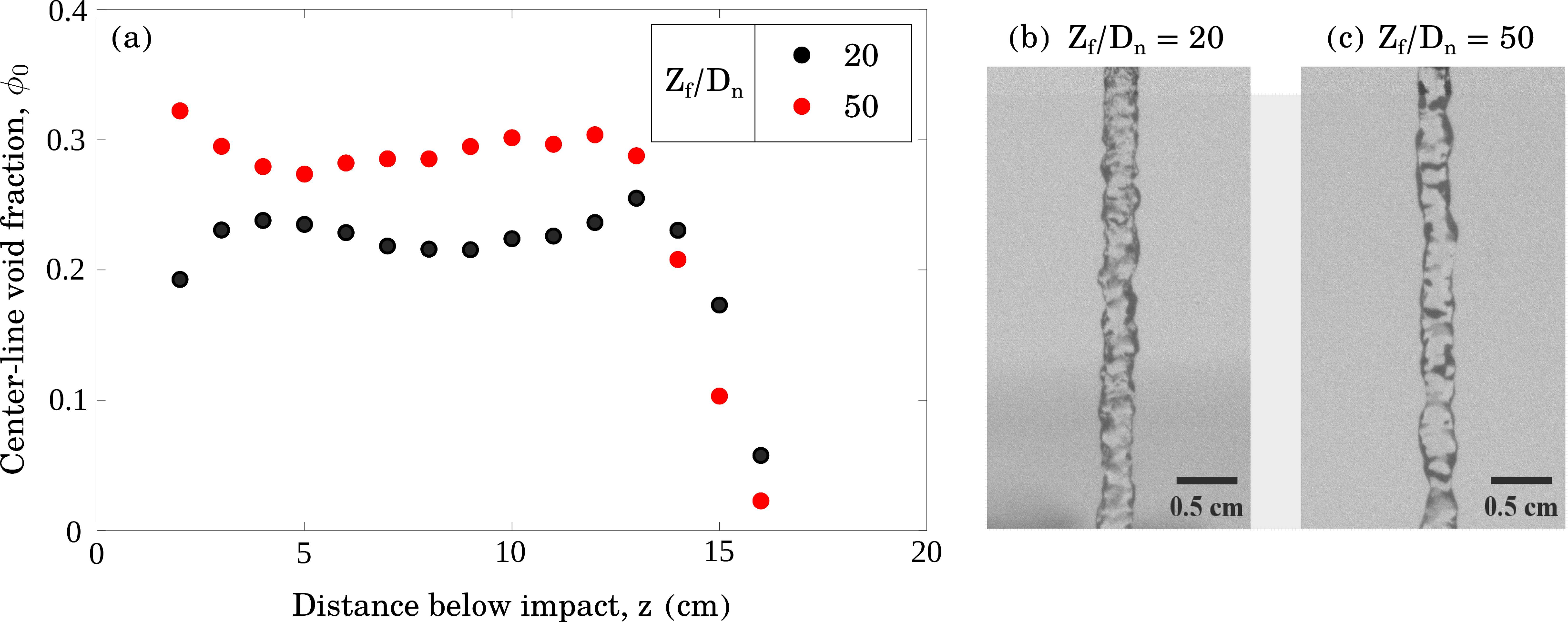,width=1\textwidth,keepaspectratio=true}
	\end{center}
  \caption{(a) Axial {evolution} of void fraction $\phi_0 (z)$ for two different $Z_f/D_n$ at a fixed jet diameter $D_n = 2.7$~mm and $\sqrt{V_iV_n}D_n = 0.016\pm 0.0002$~m$^2$/s. {Instantaneous} images of the {freely-falling} jet just before impact {are given in (b) and (c).}}
\label{fig:phi0}
\end{figure}

Figure  \ref{fig:phi0}(a) {presents air to water volume fraction at the jet axis, $\phi(r = 0, z)$, denoted here as $\phi_0$. These measurements correspond to} the bubble cloud generated by the $D_n = 2.7$~mm nozzle for two distinct values of jet fall-height to diameter ratio $Z_f/D_n$ while maintaining nearly identical impact momentum. The values of $\phi_{0}$ are nearly constant with depth $z$, until they sharply decrease at the end of the cloud, as previously reported by \cite{van1981water}, \cite{mckeogh1981air}, \cite{chanson2003air}, and \cite{hoque2008air}. 
When $Z_f/D_n$ is increased, a similar axial profile is {observed}, but $\phi_{0}$  is increased by {almost} $26$~\%. 

The jet morphology just before impact is illustrated in Fig. \ref{fig:phi0}(b) and (c) for $Z_f/D_n = 20$ and $Z_f/D_n = 50$, and it shows a slightly larger amplitude of the jet undulations for the larger $Z_f/D_n$ case. {Using the algorithm introduced in Section 2, we measure that the undulations on the jet surface exhibit a roughness of $\epsilon = 170$ $\mu$m for Figure 3b and $\epsilon = 225$ $\mu$m for Figure 3c.} As discussed by \cite{mckeogh1981air} and \cite{sene1988air}, we expect the {air entrainment rate} {and therefore the void fraction} to be proportional to the volume enclosed by the {corrugations observed} on the jet surface. The increase in the size of these undulations is therefore certainly what leads to the increase in void fraction observed in figure \ref{fig:phi0}(a). Supplementary movies 1 and 2 provide additional visualizations of the jets of figure  \ref{fig:phi0}(b) and (c) respectively (frame rate of 300 images/s).
These two conditions are indicated by labels I and II in Fig. \ref{fig:HVsSqrtViVnDn}. They are both situated in the inertia-dominated regime, where {buoyancy effects are still negligible.} As a result, the effect of void fraction on $H$ is negligible, even though $\phi$ increases. 

 \subsubsection{Radial measurements}
 \label{subsubsec:phi_radial}
\begin{figure}
	\begin{center}
		\epsfig{file=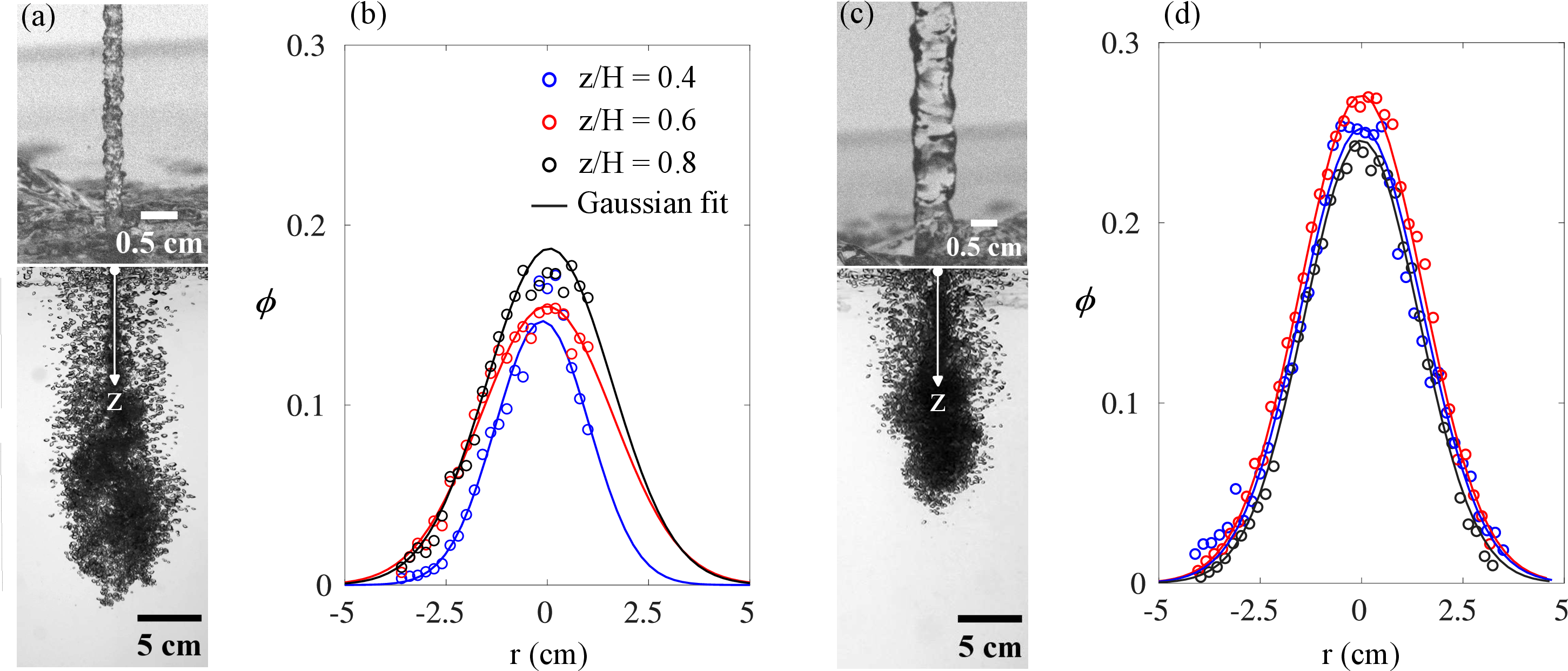,width=1\textwidth,keepaspectratio=true}
	\end{center}
  \caption{{(a) and (c) present instantaneous pictures of the falling jet just before impact for  $D_n = 2.7$ and $8$~mm, respectively, at constant $\sqrt{V_nV_i}D_n = 0.028$~m$^2$/s and $Z_f/D_n = 20$. Corresponding bubble clouds are shown as well.} (b) and (d) illustrate the radial variation of $\phi(r, z)$ at three different depths in the bubble cloud, for the $2.7$~mm and $8$~mm jets, respectively.}
\label{fig:phi_dia_effect}
\end{figure}

{Radial profiles of the void fraction are measured for three nozzle diameters}, and two heights of fall, {at} a constant impact momentum. {The radial profiles are taken at depths larger than $4D_i$, in order to ensure they are in the developed region \citep{ervine1987behaviour}}. Fig. \ref{fig:phi_dia_effect}(a) shows the bubble cloud generated by a jet of nozzle diameter $D_n=2.7$~mm. In comparison, the bubble cloud depth issued from the larger $D_n = 8$~mm, as shown in Fig. \ref{fig:phi_dia_effect}(c), is significantly smaller (see also, labels III and IV in Fig. \ref{fig:HVsSqrtViVnDn}). The visualizations of figure \ref{fig:phi_dia_effect}(a) and (c) are illustrated more extensively in supplementary movies 3 to 6. The  void fraction profiles for the 2.7 and {8}~mm injectors are shown in Fig. \ref{fig:phi_dia_effect}(b) and (d), respectively. They are measured for three different depths. Dispersion in $\phi$ profiles is relatively small in the developed region. These profiles can be well approximated using a Gaussian distribution (solid lines) as previously reported by \cite{van1981water}. This finding is similar to the case when an air jet is injected into the pool \citep{kobus1968analysis,freire2002bubble}. 
bis{When our data is compared with fits based on air/bubble diffusivity and air-to-water volume flux ratio, as in \citet{cummings1997air}, the best fit was found for the air-to-water entrainment ratio of O(10)! This unrealistic value results perhaps from their constant-diffusivity-assumption for advection-diffusion process in the two-phase mixing layer. In our case, and as described by \citet{clanet1997depth}, the bubbles are convected by the large-scale eddies dominating the evolution of the submerged jet.}


Note that the maximum void fraction $\phi(r = 0, z)$ is significantly smaller for the {thinner} jet (up to $18$~\% along the axis, Fig.\ref{fig:phi_dia_effect}b) when compared with that of the larger jet (up to 27 \% in Fig. \ref{fig:phi_dia_effect}d). {The impact of the nozzle diameter on the maximum void fraction has already been evidenced by \citet{chanson2004physical}}. The larger void fraction, and hence buoyancy force, is consistent with the observation that the bubble cloud is smaller for the larger diameter. {Furthermore, this increase in void fraction is very likely caused by the larger length scales associated with the corrugation and its amplitude occurring on the jet surface when jet diameter $D_n$ is increased} at a constant $Z_f/D_n = 20$ and constant impact momentum \citep{liu2022consistent,sene1988air}. {Measurements of $\epsilon$ show that the roughness goes from 225 $\mu$m for Fig. 4b to 445 $\mu$m for figure 4d}. 
These points are indicated by labels III and IV in Fig. \ref{fig:HVsSqrtViVnDn}. It is clear that III belongs to the momentum-dominated regime, while IV belongs to the buoyancy-dominated regime.


\begin{figure}
	\begin{center}
		\epsfig{file=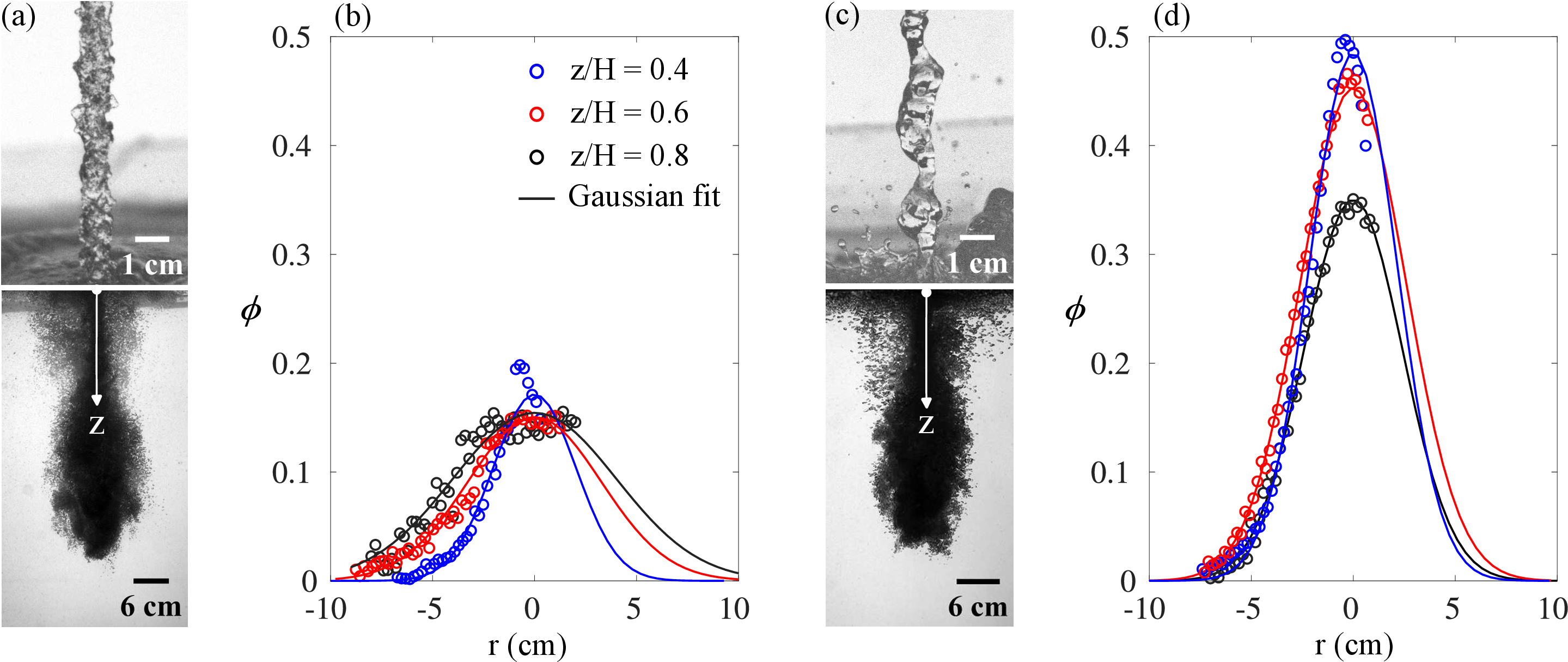,width=1\textwidth,keepaspectratio=true}
	\end{center}
  \caption{Images of $D_n$ = 8 mm jet just before impact for (a) $Z_f/D_n$ = 20  and (c) 100, for a constant $\sqrt{V_nV_i}D_n = 0.075$~m$^2$/s. Corresponding bubble clouds are shown as well.  (b) and (d) illustrate the radial variation of $\phi$ at various depths in the bubble clouds for $Z_f/D_n$ = 20 and 100 respectively.}
\label{fig:phi_fall_effect}
\end{figure}

\begin{figure}
     \centering
     \includegraphics[scale=0.45]{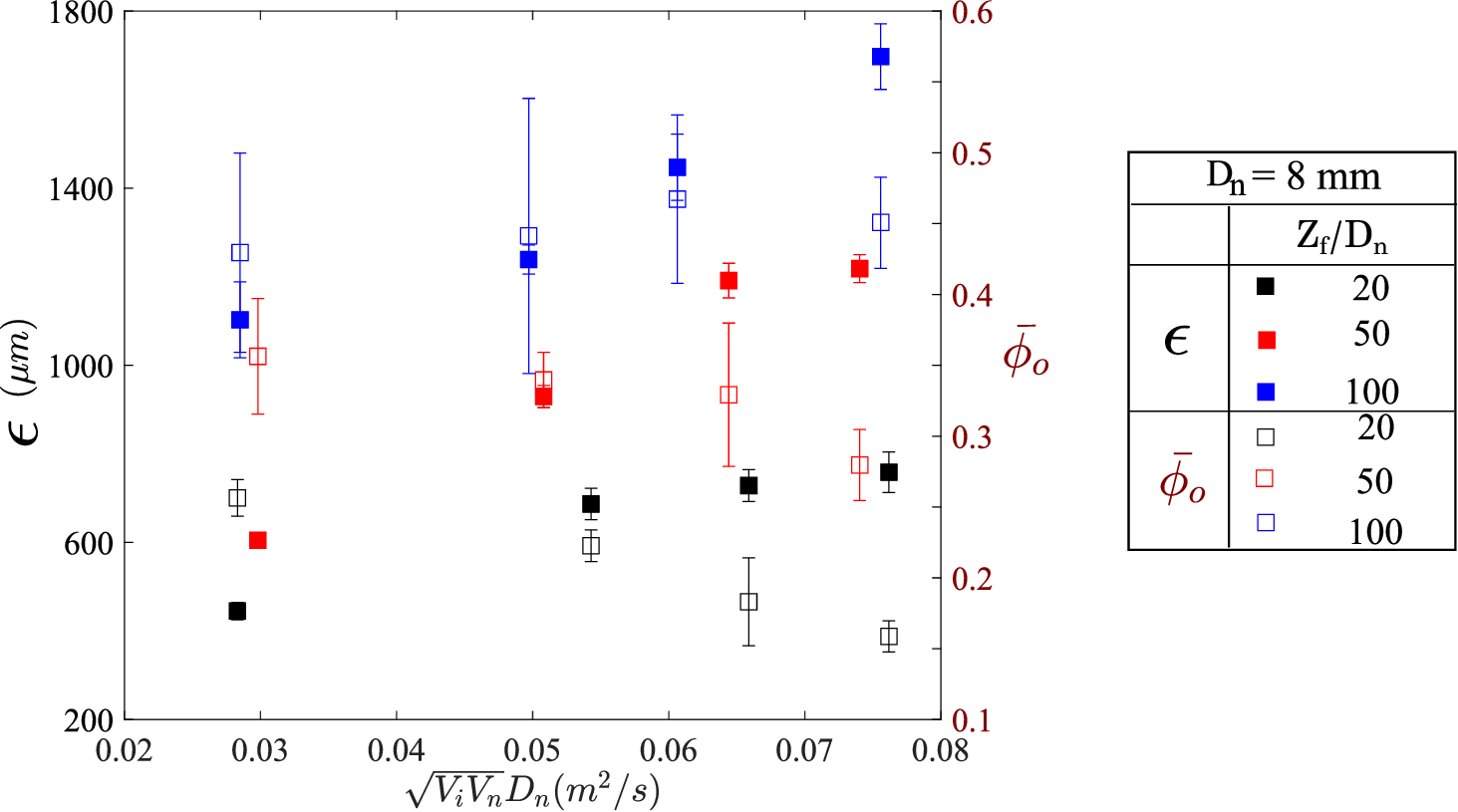}
     \caption{{Variation of $\epsilon$ and $\bar{\phi}_{0}$ with $\sqrt{V_iV_n}D_n$ for $D_n = 8$~mm, where $\bar{\phi}_{0}$ is the average over depth of the maximum void fraction measured along the axis. Both $\epsilon$ and $\bar{\phi}_{0}$ increase when $Z_f$ is increased for a constant jet momentum.}}
     \label{Figureepsilon}
 \end{figure}

Figure \ref{fig:phi_fall_effect} illustrates the influence of a change of the dimensionless jet fall height $Z_f/D_n$, for a constant impact momentum, and constant diameter $D_n = 8$~mm. {Photographs of the} jet before impact show that the jet issued at $Z_f/D_n = 100$ (Fig. \ref{fig:phi_fall_effect}c) is more {corrugated} than that issued at $Z_f/D_n = 20$ (Fig. \ref{fig:phi_fall_effect}a), for the same $\sqrt{V_iV_n}D_n=0.075$~m$^2$/s. {This is corroborated by the data of figure \ref{Figureepsilon}, which shows the variation of roughness of the jet at impact ($\epsilon$), and the variation of $\bar{\phi}_{0}$, the average over depth of the maximum void fraction measured along the axis, as a function of $\sqrt{V_iV_n}D_n$ for different $Z_f$. The data show a strong increase in $\epsilon$ and $\bar{\phi}_{0}$ as $Z_f$ is increased for a constant momentum.}
\begin{figure}
	\begin{center}
		\epsfig{file=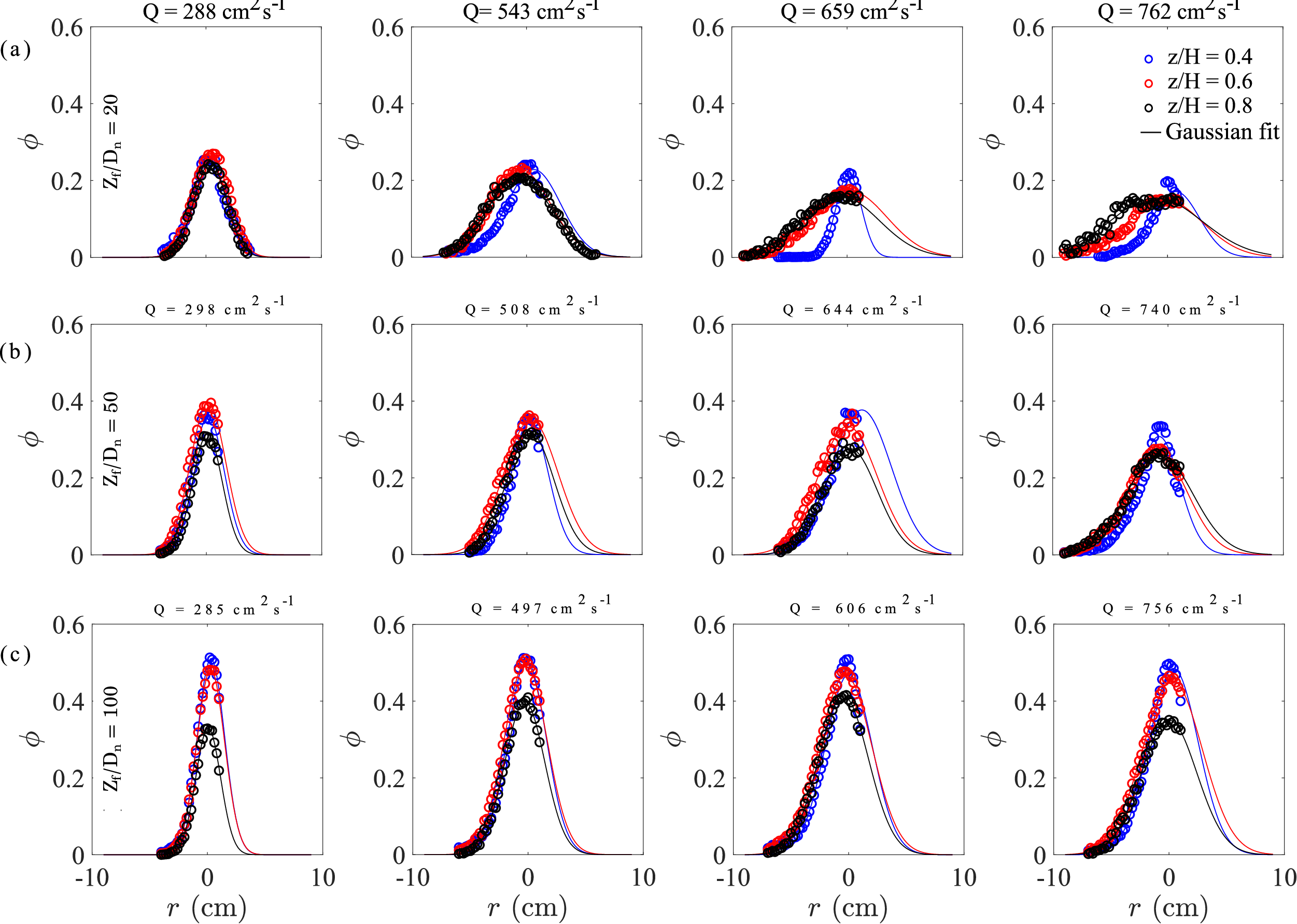,width=1\textwidth,keepaspectratio=true}
	\end{center}
  \caption{Measured profiles of air/water void fraction for the nozzle diameter $D_n = 8$~mm and different fall-height (a) $Z_f/D_n = 20$, (b) $Z_f/D_n = 50$, and (c) $Z_f/D_n = 100$  at various Q = $\sqrt{V_iV_n}D_n$. Gaussian fits are represented by continuous lines.  }
  
\label{fig:phi_variation}
\end{figure}
The visualizations of figure \ref{fig:phi_fall_effect}(a) and (c) are illustrated more extensively in supplementary movies 7 to 10. The bubble cloud generated below the surface is smaller for the larger $Z_f/D_n$. Again, this is likely due to the fact that this jet captures more air within its cavities {during free-fall}, leading to a higher $\phi$. This is confirmed by the radial profiles given in figures \ref{fig:phi_fall_effect}(b) and (d), which show that $\phi_0$ for $Z_f/D_n  = 100$ is approximately $3.2$ times that for $Z_f/D_n  = 20$. {Note that these cases are referred to as labels V and VI, respectively, in figure \ref{fig:HVsSqrtViVnDn}. They both correspond to the buoyancy-dominated regime.}
{In addition, the inset displayed in Figure 2 illustrates a zoom view of  the change in $H$ at higher $\sqrt{V_n V_i } D_n$, revealing that an increase in $Z_f /D_n$ results in a reduction of $\textit{H}$, with a decrease of approximately 20$\%$ observed when $Z_f /D_n$ is raised from 20 to 100}. This ascertains that the observed dispersion for the cloud depth in figure \ref{fig:HVsSqrtViVnDn}, and also in previous investigations provided in \citet{guyot2020penetration} for this regime at higher impact momentum and larger diameter, is due to differences in jet fall height.

{Finally, general trends of local void fraction profiles are depicted in figure \ref{fig:phi_variation} for the case of $D_n = 8$~mm. Each figure presents data at different vertical locations $z$ underneath the surface, normalized with respect to the cloud depth $H$. In all cases shown here, data fits reasonably well with a Gaussian profile $\phi = \phi_0(z) \mbox{e}^{-\left( r^2/\sigma^2(z) \right)}$ whose peak and width do not vary much as $z/H$ is changed. The first row in figure \ref{fig:phi_variation} corresponds to data for the shortest jet fall height to diameter ratio $Z_f/D_n = 20$ at various $\sqrt{V_iV_n}D_n$, a measure of the jet impact momentum up to a factor of liquid density. As $\sqrt{V_iV_n}D_n$ increases the centreline peak void fraction $\phi_0$ slightly decreases while the width of the best-fitted Gaussian profiles increases. On the other hand, at a given $\sqrt{V_iV_n}D_n$, the peak void fraction $\phi_0$ is almost doubled when the jet's fall height is quintupled. In the next section, these observations on the air/water volume fraction in the biphasic region are further developed to properly capture the effect of buoyancy on the bubble cloud depth.} 

\subsection{Prediction of bubble cloud depth}
\label{subsec:prediction}

{\citet{guyot2020penetration} proposed a prediction for the bubble cloud depth $H$ via Eq. (\ref{eq:guyot}) by assuming a uniform void fraction within the bubble cloud and a constant cone angle $\gamma$, based on single-phase turbulent jets. As mentioned just before, the void fraction in the biphasic jet underneath the free surface can be well-approximated by a Gaussian profile.} These profiles are characterised by a maximum void fraction $\phi_0(z)$ on the axis, and a variance $\sigma(z)$. It is {then} easy to show that the buoyancy force on a slice of bubble cloud of thickness $dz$ at a depth $z$ is $dF_b = \phi_{0}(z) \rho g \pi\sigma(z)^2 dz$. We take $\phi_0(z)$ to be equal to its average value $\bar{\phi}_{0}$, since the peak void fraction $\phi_0(z)$ exhibits little variation with depth $z$, as already shown in figures \ref{fig:phi0} to \ref{fig:phi_fall_effect}. It now remains to provide $\sigma(z)$ in order to integrate this buoyancy force over $z$.

{Indeed, it is expected that} $\sigma(z)$ increases linearly with depth as the bubble cloud widens up as a cone, as evidenced by visualization {and previous observations \citep{suciu1976gas, mckeogh1981air, clanet1997depth, guyot2020penetration}. We propose to measure the angle of this cone based on the experimental void fraction profiles. However, rising bubbles can be detected outside the conical jet region by the optical probe. This may lead to overestimating the width of the void fraction profiles, and hence the resulting buoyancy force on the bubble cloud. This is particularly true for the measurements at shallower depths $z < 0.5H$, a problem also encountered by \cite{van1981water}. In order to circumvent this difficulty, the conical jet hypothesis is maintained in accordance with previous authors so that $\sigma(z) = D_i/2+ z \tan \gamma_{0}$, where the cone angle $\gamma_0$ is computed as $\tan \gamma_{0} = \left(\sigma(\tilde{z}) - D_i/2 \right) /\tilde{z}$ at a chosen reference depth $\tilde{z} = 0.8 H$. The latter depth was chosen to avoid both rising bubbles and the steep decrease in bubble void fraction in the neighbourhood of the cloud's tail at $z = H$.}
 

Thereby, the net buoyancy force $F_b$ on the bubble cloud {of depth $H_b$} can then be expressed as
\begin{equation}
	F_b =  \bar{\phi}_{0} \rho g {\pi}\left( {\frac{1}{3}} H_b^3 \tan^2 \gamma_{0} + {\dfrac{D_i}{2}} H_b^2 \tan \gamma_{0} + {\dfrac{D_i^2}{4}} H_b \right),
\label{eq:buoyancy}
\end{equation}
{which is identical to the volume term in Eq. (\ref{eq:guyot}), if $\bar{\phi} = \bar{\phi}_{0}$ and $\gamma = \gamma_{0}$. The above expression indicates} that the buoyancy force on a conical biphasic jet exhibiting a Gaussian void fraction profile can be interpreted as the buoyancy force exerted on an equivalent biphasic jet with a constant void fraction $\bar{\phi}$ equal to the \textit{maximum} void fraction $\bar{\phi}_{0}$ of the Gaussian profile, and with a width defined by the variance of the Gaussian profile at some reference depth, here chosen at $z = 0.8H$. This buoyancy force $F_b$ can now be injected into the momentum balance (equation \ref{eq:guyot}) in order to solve for $H_b$. The value of $\bar{\phi}$ in the outgoing momentum term {can be} taken equal to $\bar{\phi}_{0}$ for the sake of simplicity.
\begin{figure}
	\begin{center}
		\epsfig{file=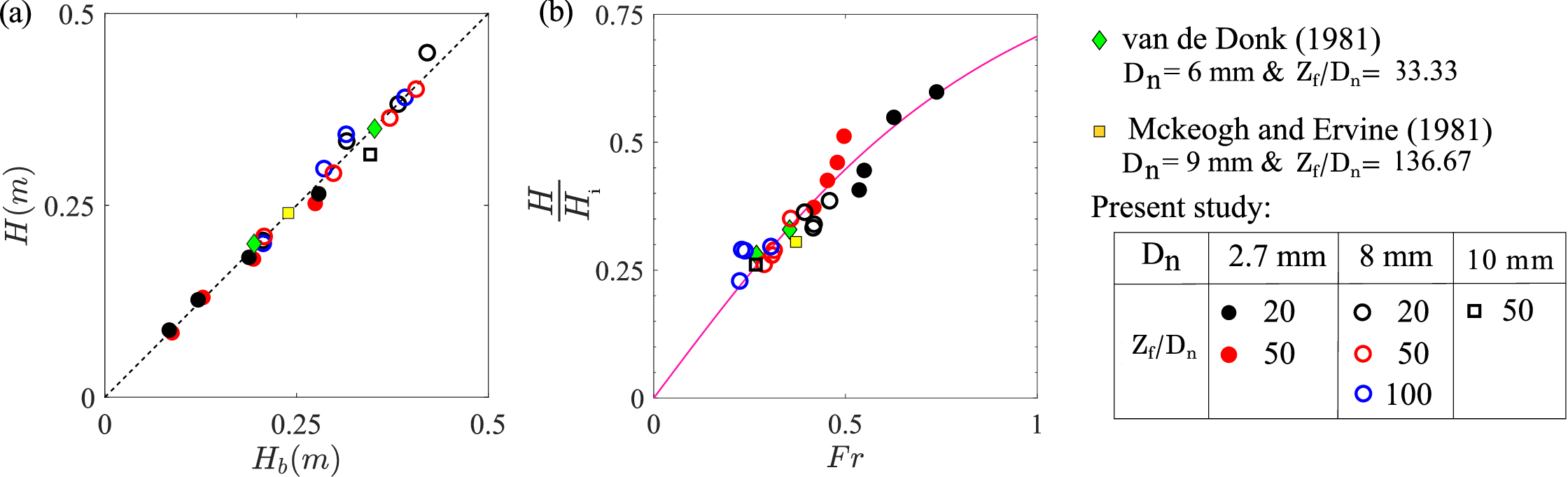,width=1\textwidth,keepaspectratio=true}
	\end{center}
   \caption{(a)  Comparison of experimental depth $H$ with {those} predicted by the model $H_b$ and (b) Normalized depth $H/H_i$ at various $Fr$ for injectors from current and past studies {whose void fraction is known}. Eq. (\ref{eq:HinFroude}) is represented by the continuous line.}
\label{fig:froude}
\end{figure}

\begin{figure}
	\begin{center}
		\epsfig{file=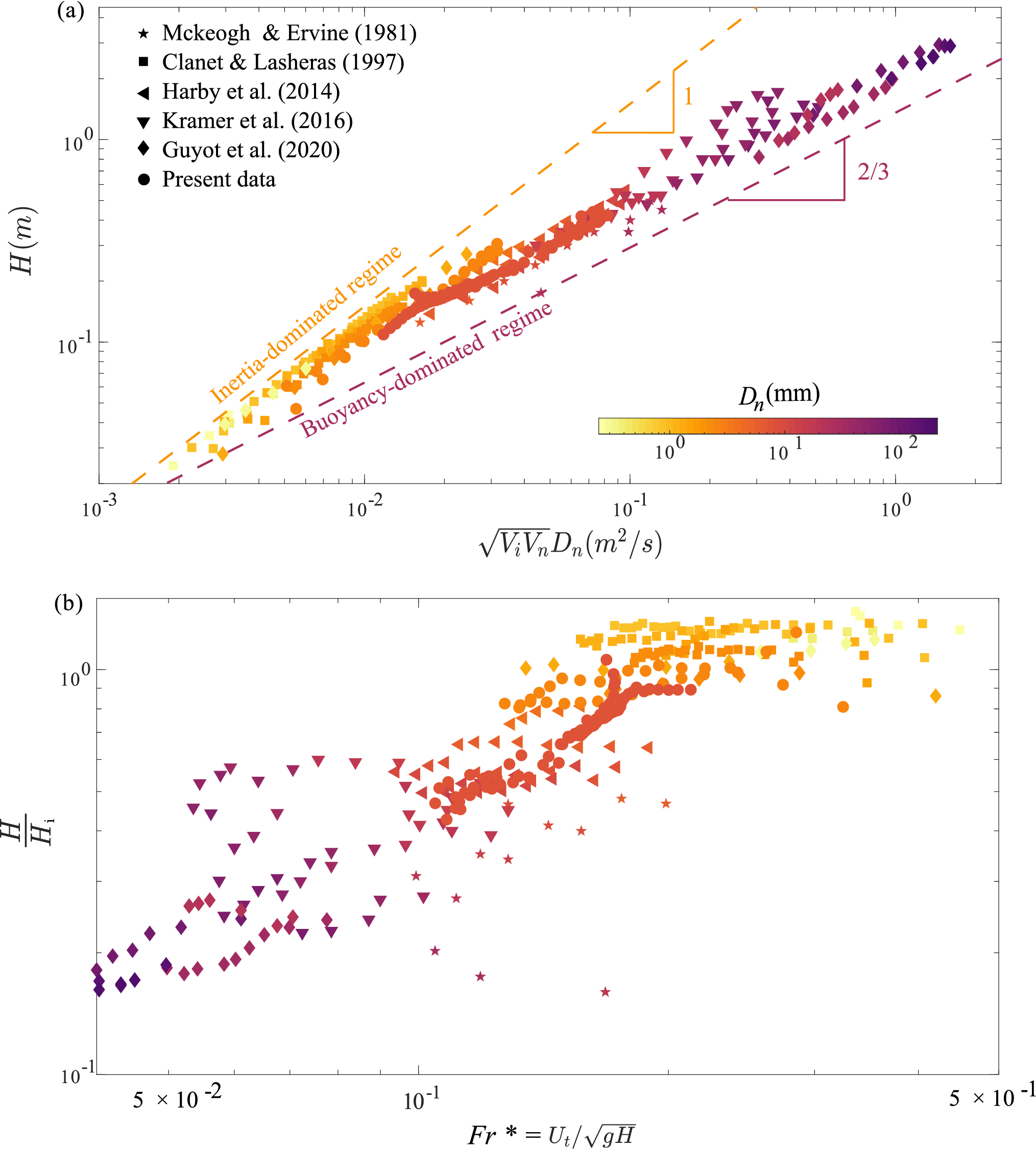,width=1\textwidth,keepaspectratio=true}
	\end{center}
   \caption{(a) Bubble cloud depth ($H$) data from previous investigations are presented here as a function of $\sqrt{V_iV_n}D_n$, a measure of the impact momentum, to illustrate the two distinct dynamical regimes identified in our study over a wide range of scales.  (b) When void fraction is not known, $Fr^*$ can still sort past experiments into inertia or buoyancy-dominated regimes, even though more dispersion is observed. }
\label{fig:froude2}
\end{figure}

Experimental depths $H$ from the present study and predictions $H_b$ are compared in Fig. \ref{fig:froude}(a) for the different nozzle sizes and jet fall heights. A good agreement is found between the model and measurements for both the inertia-controlled bubble clouds ($D_n = 2.7$~mm) and the buoyancy-controlled ($D_n = 8$~mm and 10~mm) ones. A measurement for nozzle diameter $D_n=10$~mm (same length-to-diameter ratio of 50), $V_n=5.5$~m/s and $Z_f/D_n=50$ has been included in the data of this figure.   
For comparison, two previous investigations, namely, those conducted by \cite{van1981water} and \cite{mckeogh1981air} are presented in Fig. \ref{fig:froude}(a) as well. They provided both $H$ measurements and radial profiles of $\phi$ for jets with diameters of $D_n = 6$ and $9$~mm. The calculated $H_b$ values based on their measurements show a very good consistency with the model.

{In what follows, the question of the role of void fraction in the transition from the inertia-dominated to the buoyancy-dominated regime is discussed. This transition is expected to be dominated by the ratio of the outgoing momentum to buoyancy force, respectively second and third terms in Eq. (\ref{eq:guyot}). The square root of the ratio of these terms defines a dimensionless grouping similar to a Froude number, referred to as $Fr = \sqrt{3\left( 1-\bar{\phi}_{0} \right)/\bar{\phi}_{0}} U_t/\sqrt{g H_b}$. In fact, when the nozzle diameter is small compared to the cloud width $2 H_b \tan \gamma_0$,
the cubic polynomial for $H_b$ in Eq. (\ref{eq:guyot}) can be rewritten in terms of $Fr$ as
\begin{equation}
	\dfrac{H_b}{H_i} = \dfrac{Fr}{\sqrt{1 + Fr^2}} + \mathcal{O}\left( \dfrac{D_i}{2 H_b \tan \gamma_0} Fr^2\right),
	\label{eq:HinFroude}
\end{equation}
where $H_i$ is the cloud depth when the buoyancy force is absent, as given by Eq. (\ref{eq:clanet}).} When $Fr$ is large, we expect the cloud to be in the inertia-dominated regime, and hence $H_b$ to be close to $H_i$, the simple prediction by \citet{clanet1997depth}. When on the contrary $Fr$ is small, $H$ is expected to be controlled by buoyancy, and hence to be significantly smaller than $H_i$. Figure \ref{fig:froude}(b) illustrates the variation of $H/H_i$ with $Fr$ for data from Fig. \ref{fig:froude}(a).
Indeed, all experimental data points fall on a single curve given by Eq. (\ref{eq:HinFroude}), for which $H/H_i$ falls steeply when $Fr$ is decreased. For a constant velocity, using larger nozzles will lead to {bigger} bubble clouds, and so larger $H$ and smaller $Fr$, thereby favouring the buoyancy-dominated regime. Note also that the influence of jet velocity is counter-intuitive here since a larger jet velocity for a given nozzle diameter will generate a larger $H$, and therefore also favour the buoyancy-dominated regime, instead of the inertial regime. This is because the relevant velocity scale in $Fr$ is the terminal bubble velocity $U_t$, which is a constant, and not the jet impact velocity. {Note that expression (\ref{eq:HinFroude}) is obtained from Eq. (\ref{eq:guyot}) by assuming $\bar{\phi} = \bar{\phi}_{0}$ and $\gamma = \gamma_{0}$. This choice simplifies the expression of the Froude number, but obviously underestimates the jet angle and mean void fraction in the inertial contribution. This is partly the reason why $H/H_i$ does not reach the limit $H/H_i = 1$ in figure \ref{fig:froude}(b), even for data points which corresponded to the inertia-dominated regime in Fig. \ref{fig:HVsSqrtViVnDn}.} Furthermore, previous authors have proposed to add a hemispherical dome to the truncated cone modelling the bubble cloud, as in figure \ref{fig:schematic} \citep{clanet1997depth, guyot2020penetration}. This small correction would introduce a factor $1 + \tan \gamma_0$ (of the order of 10 \%) to both the modelled heights $H_i$ and $H_b$, but is not considered in the present data for the sake of simplicity. 

{For most studies for which bubble cloud depths data are available, the void fraction has not been measured concomitantly. We plot on Fig. \ref{fig:froude2}(a) $H$ as a function $\sqrt{V_iV_n}D_n$, square-root of the impact momentum over liquid density $\rho$, for five past studies covering a wide range of scales, plus the present data. The scaling laws observed in figure \ref{fig:HVsSqrtViVnDn}, for the inertia-dominated and buoyancy-dominated regimes, can be observed in the limit of low momentum and large momentum respectively.} 
The void fraction is not known for these past works, but we can still estimate the bubble cloud Froude number as $Fr^* = U_t/\sqrt{gH}$. Figure \ref{fig:froude2}(b) shows that when $H/H_i$ is plotted as a function of $Fr^*$ for several past studies, the cloud sizes are clearly sorted in two regimes: (i) One for larger $Fr^*>0.2$ for which $H/H_i$ is close to one, and for which there is little dispersion, and (ii) A second regime for $Fr^*<0.2$ where $H/H_i$ varies between $0.15$ and $0.5$, and for which a significant dispersion is obtained. This dispersion is very likely due to the role of the void fraction, which has been omitted in $Fr^*$. These observations further ascertain that $Fr$ is a good parameter to monitor the transition between the inertia-dominated and the buoyancy-dominated regime.


\subsection{Equivalent bubble cloud void fraction $\phi_{b}$}
 
Whereas Guyot et al (2020) took a uniform void fraction within the cloud to express the momentum balance, $\phi$ profiles were observed here to follow a Gaussian distribution in the radial direction. In this section, an \textit{equivalent} constant void fraction $\phi_{b}$ is proposed in order to provide a bulk quantity for a bubble cloud and to discuss its variations with the {jet fall height} and impact momentum. {At first, for a fixed depth $z$, the edge of this constant-void-fraction-cone of bubble cloud is defined by admitting that its radius $R_{b}(z)$, beyond which the void fraction becomes zero, is the radius at which $80$\% of the surface-integrated void fraction is attained based on a Gaussian profile $\phi = \bar{\phi}_0(z) e^{-r^2/\sigma(z)^2}$. This simply gives the relation $R_{b}(z) = \sigma(z) \sqrt{\log{5}} \simeq 1.27 \sigma$. Thereafter, in order to avoid over-estimation of the bubble cloud volume due to rising bubbles, the truncated cone angle $\gamma_{b}$ for this constant-void-fraction cloud is taken as $\tan \gamma_{b} = (\tilde{R}_{b}- D_i/2)/\tilde{z}$ where $\tilde{R}_{b} \simeq 1.27 \sigma(\tilde{z})$ is this cloud's radius at the reference depth $\tilde{z} = 0.8H$. As already mentioned in section \ref{subsec:prediction}, the influence of rising bubbles is minimal at the reference depth $\tilde{z} = 0.8H$, measured close to the bottom of the cloud while the \textit{peak} void fraction ${\phi}_0$ is still comparable to those obtained in the bulk.} Finally, the \textit{equivalent} void fraction $\phi_b$ is then computed such that the net buoyancy force is the same on both the uniform-void-fraction cloud and the Gaussian-void-fraction cloud.

\begin{figure}
	\begin{center}
		\epsfig{file=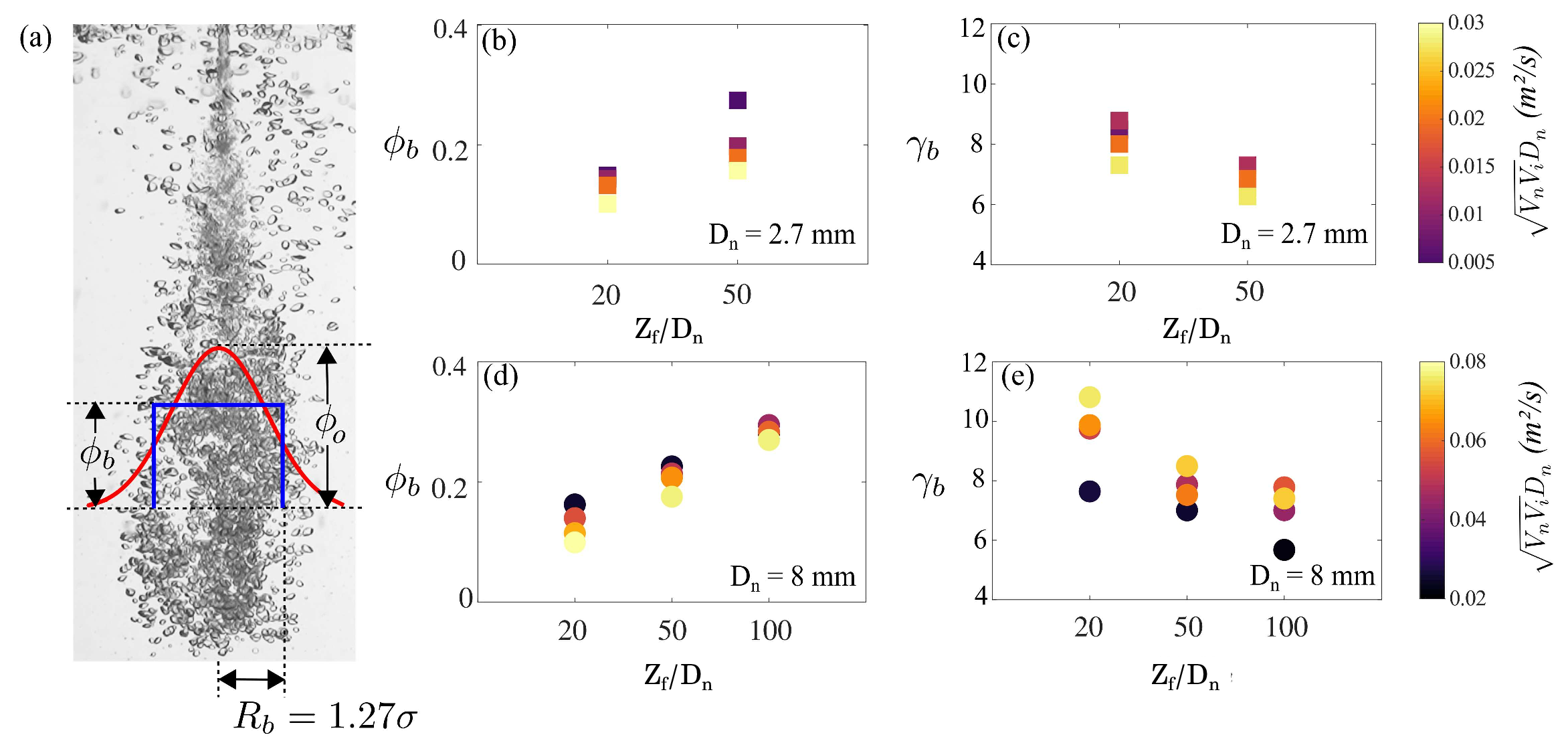,width=1\textwidth,keepaspectratio=true}
	\end{center}
  \caption{(a) Definition of $\Phi_{b}$ and $R_{b}$. (b) Variation of $\Phi_{b}$ and $\gamma_{b}$ with $Z_f/D_n$ for the two injector sizes at various impact momentum. Two color bars are provided for the two ranges of $\sqrt{V_iV_n}D_n$, corresponding to both $D_n$ values.}
\label{fig:equiv_phi}
\end{figure}

Figures \ref{fig:equiv_phi}(b) and (d) depict the evolution of this \textit{equivalent} void fraction $\phi_{b}$ as a function of $Z_f/D_n$, for different values of impact momentum (represented by the colour) for the two injectors. Experimental results show that $\phi_{b}$ increases steeply for both injectors as $Z_f/D_n$ increases, as already illustrated in figure \ref{fig:phi_fall_effect} for a particular case. As mentioned in subsection \ref{subsubsec:phi_radial}, this is likely due to the jet's capacity to entrain more air once perturbations have grown larger on its surface for higher fall height $Z_f$ \citep{sene1988air}. These plots also show that $\phi_{b}$ decreases for a constant $Z_f/D_n$  as the impact momentum is increased, in particular for low $Z_f/D_n$. 

Figure \ref{fig:equiv_phi}(c) and (e) illustrate the variations of the cone angle $\gamma_{b}$ of the \textit{equivalent} constant-$\phi$ jet with regards to $Z_f/D_n$, for different $\sqrt{V_iV_n}D_n$. Results for both nozzles show that $\gamma_{b}$ decreases with $Z_f/D_n$ for a constant impact momentum. This change in the bubble cloud shape may be related to the strong increase in the amplitude of surface perturbations at larger $Z_f$ and to the larger void fractions observed for these conditions ($\phi_0$ up to $50$\% for $Z_f/D_n = 100$, compared to $20$\% for $Z_f/D_n = 20$). The simple assumptions made regarding the shape of the bubble cloud,  which are valid at moderate void fractions, probably do not hold anymore in such conditions.



\section{Conclusion}
Experiments {on circular plunging jets} have been carried out to characterize the two regimes for bubble cloud formation, namely the inertia-dominated and the buoyancy-dominated regimes.  Void fraction profiles were measured in the developed region of the cloud with novel optical probes {for nozzle diameter, $D_n$ = 2.7, 8 and 10 mm.} They can be well approximated by Gaussian distributions. When these {measurements are incorporated into the} momentum balance model, taking into account only the buoyancy force, it predicts the experiments very well for both current and past studies. {Even though the void fraction measurements have only been carried out for nozzle sizes smaller than 1 cm and bubble cloud sizes smaller than 50 cm, it is expected that this result  still holds at larger scales, as already proposed by \citet{guyot2020penetration}. This will have to be verified with void fraction measurements in such metric scale systems in future works. }

The threshold between the inertia and buoyancy regimes is shown to be dependent on a \textit{characteristic} Froude number $Fr$ built with the terminal bubble velocity $U_t$, the cloud depth $H$, and the void fraction within the bubble cloud. For large values of this number, $H$ can be estimated with the simple model of \cite{clanet1997depth}, while for low values of $Fr$, it is buoyancy that controls the bubble cloud size. {The fact that a single equation can predict bubble cloud size independent of scale and of the wide zoology of jet dynamics provided the void fraction is known is the major result of this work.}

A constant \textit{equivalent} void fraction $\phi_{b}$ is also introduced based on the integration of the Gaussian void fraction profiles. The variations of $\phi_{b}$ and of the corresponding jet cone angle $\gamma_{b}$ with the impact momentum and the jet fall-height to diameter ratio $Z_f/D_n$ are discussed. 
This void fraction $\phi_{b}$ is expected to be a function of jet diameter, the height of fall, and velocity \citep{wang2018intrusive}. Data show that an increase in the height of fall $Z_f$ leads to a significant increase in the void fraction within the cloud, which in turn leads to a decrease in the bubble cloud size. The modelling of $\phi_{b}$ as a function of jet dynamics, {which is necessary for practical purposes if one wishes to predict $H$ solely as a function of input parameters,} is the question future works will have to address.

\medskip

\textbf{Acknowledgements.} We acknowledge the technical support of G. Geniquet, S. Martinez and A. Buridon on the experimental set-up. We also thank A2 Photonic Sensors for the support with the optical probes. This research was funded by the French Agence Nationale de la Recherche ANR under grant no. JETPLUME ANR-21-CE05-0029.

\medskip

\textbf{Funding.} -  This research was funded by the French Agence Nationale de la Recherche ANR under grant no. JETPLUME ANR-21-CE05-0029.

\medskip

\textbf{Declaration of interest.} The authors report no conflict of interest.

\medskip

\textbf{Author ORCIDs.} Narendra Dev, \href{https://orcid.org/0009-0007-4626-3585}{0009-0007-4626-3585}; J. John Soundar Jerome, \href{https://orcid.org/0000-0003-2148-9434}{0000-0003-2148-9434}; H. Scolan, \href{https://orcid.org/0000-0003-2739-2297}{0000-0003-2739-2297}; J.-P. Matas, \href{https://orcid.org/0000-0003-0708-1619}{0000-0003-0708-1619};


\begin{thebibliography}{44}
\providecommand{\natexlab}[1]{#1}
\providecommand{\url}[1]{\texttt{#1}}
\expandafter\ifx\csname urlstyle\endcsname\relax
  \providecommand{\doi}[1]{doi: #1}\else
  \providecommand{\doi}{doi: \begingroup \urlstyle{rm}\Url}\fi
\bibitem[Bertola et~al.(2018)Bertola, Wang, and Chanson]{bertola2018physical}
N.~Bertola, H.~Wang, and H.~Chanson.
\newblock A physical study of air--water flow in planar plunging water jet with
  large inflow distance.
\newblock \emph{International Journal of Multiphase Flow}, 100:\penalty0
  155--171, 2018.

\bibitem[Bi{\'n}(1993)]{bin1993gas}
Andrzej~K Bi{\'n}.
\newblock Gas entrainment by plunging liquid jets.
\newblock \emph{Chemical Engineering Science}, 48\penalty0 (21):\penalty0
  3585--3630, 1993.

\bibitem[Bonetto and Lahey(1993)]{bonetto1993ijmf}
F.~Bonetto and R.~T. Lahey.
\newblock An experimental study on air carryunder due to a plunging liquid jet.
\newblock \emph{International Journal of Multiphase Flow}, 19\penalty0
  (2):\penalty0 281--294, 1993.
\newblock ISSN 0301-9322.
\newblock \doi{https://doi.org/10.1016/0301-9322(93)90003-D}.

\bibitem[Bonetto et~al.(1994)Bonetto, Drew, and Lahey~Jr]{bonetto1994analysis}
F~Bonetto, D~Drew, and RT~Lahey~Jr.
\newblock The analysis of a plunging liquid jet -- the air entrainment process.
\newblock \emph{Chemical Engineering Communications}, 130\penalty0
  (1):\penalty0 11--29, 1994.

\bibitem[Brattberg and Chanson(1998)]{brattberg1998air}
T.~Brattberg and H.~Chanson.
\newblock Air entrapment and air bubble dispersion at two-dimensional plunging
  water jets.
\newblock \emph{Chemical engineering science}, 53\penalty0 (24):\penalty0
  4113--4127, 1998.

\bibitem[Chanson(2002)]{chanson2002hydraulics}
H.~Chanson.
\newblock \emph{Hydraulics of stepped chutes and spillways}.
\newblock CRC Press, 2002.

\bibitem[Chanson and Manasseh(2003)]{chanson2003air}
H.~Chanson and R.~Manasseh.
\newblock Air entrainment processes in a circular plunging jet: Void-fraction
  and acoustic measurements.
\newblock \emph{J. Fluids Eng.}, 125\penalty0 (5):\penalty0 910--921, 2003.

\bibitem[Chanson et~al.(2004)Chanson, Aoki, and Hoque]{chanson2004physical}
H.~Chanson, S.~Aoki, and A.~Hoque.
\newblock Physical modelling and similitude of air bubble entrainment at
  vertical circular plunging jets.
\newblock \emph{Chemical engineering science}, 59\penalty0 (4):\penalty0
  747--758, 2004.

\bibitem[Chirichella et~al.(2002)Chirichella, Gomez~Ledesma, Kiger, and
  Duncan]{chirichella2002incipient}
D.~Chirichella, R.~Gomez~Ledesma, K.~T. Kiger, and J.~H. Duncan.
\newblock Incipient air entrainment in a translating axisymmetric plunging
  laminar jet.
\newblock \emph{Physics of Fluids}, 14\penalty0 (2):\penalty0 781--790, 2002.

\bibitem[Clanet and Lasheras(1997)]{clanet1997depth}
C.~Clanet and J.~C. Lasheras.
\newblock Depth of penetration of bubbles entrained by a plunging water jet.
\newblock \emph{Physics of Fluids}, 9\penalty0 (7):\penalty0 1864--1866, 1997.

\bibitem[Cummings and Chanson(1997)]{cummings1997air}
P.~D. Cummings and H.~Chanson.
\newblock {Air Entrainment in the Developing Flow Region of Plunging
  Jets—Part 1: Theoretical Development}.
\newblock \emph{Journal of Fluids Engineering}, 119\penalty0 (3):\penalty0
  597--602, 09 1997.

\bibitem[Cummings and Chanson(1999)]{cummings1999experimental}
P.D. Cummings and H.~Chanson.
\newblock An experimental study of individual air bubble entrainment at a
  planar plunging jet.
\newblock \emph{Chemical Engineering Research and Design}, 77\penalty0
  (2):\penalty0 159--164, 1999.

\bibitem[El~Hammoumi et~al.(2002)El~Hammoumi, Achard, and
  Davoust]{el2002measurements}
M.~El~Hammoumi, J.-L. Achard, and L.~Davoust.
\newblock Measurements of air entrainment by vertical plunging liquid jets.
\newblock \emph{Experiments in fluids}, 32:\penalty0 624--638, 2002.

\bibitem[Ervine and Falvey(1987)]{ervine1987behaviour}
D.A. Ervine and H.T. Falvey.
\newblock Behaviour of turbulent water jets in the atmosphere and in plunge
  pools.
\newblock \emph{Proceedings of the Institution of Civil engineers}, 83\penalty0
  (1):\penalty0 295--314, 1987.

\bibitem[Freire et~al.(2002)Freire, Miranda, Luz, and
  Fran{\c{c}}a]{freire2002bubble}
A.~P.~S. Freire, D.~D.~E. Miranda, L.~M.~S. Luz, and G.~F.~M. Fran{\c{c}}a.
\newblock Bubble plumes and the coanda effect.
\newblock \emph{International journal of multiphase flow}, 28\penalty0
  (8):\penalty0 1293--1310, 2002.

\bibitem[Guyot et~al.(2019)Guyot, Cartellier, and Matas]{guyot2019depth}
G.~Guyot, A.~Cartellier, and J.-P. Matas.
\newblock Depth of penetration of bubbles entrained by an oscillated plunging
  water jet.
\newblock \emph{Chemical Engineering Science: X}, 2:\penalty0 100017, 2019.

\bibitem[Guyot et~al.(2020)Guyot, Cartellier, and Matas]{guyot2020penetration}
G.~Guyot, A.~Cartellier, and J.-P. Matas.
\newblock Penetration depth of a plunging jet: from microjets to cascades.
\newblock \emph{Physical Review Letters}, 124\penalty0 (19):\penalty0 194503,
  2020.

\bibitem[Guyot(2019)]{guyot2019contribution}
Gre{\'{e}}gory Guyot.
\newblock \emph{Contribution {\`a} la caract{\'e}risation des processus
  d’entra{\^\i}nement d’air dans les circuits d’am{\'e}nagements
  hydro-{\'e}lectriques}.
\newblock PhD thesis, Universit{\'e} Grenoble Alpes, France, 2019.

\bibitem[Harby et~al.(2014)Harby, Chiva, and
  Mu{\~n}oz-Cobo]{harby2014experimental}
K.~Harby, S.~Chiva, and J.L. Mu{\~n}oz-Cobo.
\newblock An experimental study on bubble entrainment and flow characteristics
  of vertical plunging water jets.
\newblock \emph{Experimental Thermal and Fluid Science}, 57:\penalty0 207--220,
  2014.

\bibitem[Hoque and Aoki(2008)]{hoque2008air}
A.~Hoque and S.~Aoki.
\newblock Air entrainment and associated energy dissipation in steady and
  unsteady plunging jets at free surface.
\newblock \emph{Applied Ocean Research}, 30\penalty0 (1):\penalty0 37--45,
  2008.

\bibitem[Horn and Thring(1956)]{horn1956angle}
G.~Horn and M.W. Thring.
\newblock Angle of spread of free jets.
\newblock \emph{Nature}, 178\penalty0 (4526):\penalty0 205--206, 1956.

\bibitem[Kapur et~al.(1985)Kapur, Sahoo, and
  Wong]{Kapur_CompVision1985thresholding}
J.N. Kapur, P.K. Sahoo, and A.K.C. Wong.
\newblock A new method for gray-level picture thresholding using the entropy of
  the histogram.
\newblock \emph{Computer Vision, Graphics, and Image Processing}, 29\penalty0
  (3):\penalty0 273--285, 1985.
\newblock ISSN 0734-189X.
\newblock \doi{https://doi.org/10.1016/0734-189X(85)90125-2}.

\bibitem[Kiger and Duncan(2012)]{kiger2012air}
K.~T. Kiger and J.~H. Duncan.
\newblock Air-entrainment mechanisms in plunging jets and breaking waves.
\newblock \emph{Annual Review of Fluid Mechanics}, 44:\penalty0 563--596, 2012.

\bibitem[Kirillov et~al.(2023)Kirillov, Mintun, Ravi, Mao, Rolland, Gustafson,
  Xiao, Whitehead, Berg, Lo, et~al.]{kirillov2023segment}
A.~Kirillov, E.~Mintun, N.~Ravi, H.~Mao, C.~Rolland, L.~Gustafson, T.~Xiao,
  S.~Whitehead, A.~C Berg, W.-Y. Lo, et~al.
\newblock Segment anything.
\newblock \emph{arXiv preprint arXiv:2304.02643}, 2023.

\bibitem[Kobus(1968)]{kobus1968analysis}
H.~E. Kobus.
\newblock Analysis of the flow induced by air-bubble systems.
\newblock In \emph{Coastal Engineering 1968}, pages 1016--1031. 1968.

\bibitem[Kramer et~al.(2016)Kramer, Wieprecht, and
  Terheiden]{kramer2016penetration}
M.~Kramer, S.~Wieprecht, and K.~Terheiden.
\newblock Penetration depth of plunging liquid jets--a data driven modelling
  approach.
\newblock \emph{Experimental Thermal and Fluid Science}, 76:\penalty0 109--117,
  2016.

\bibitem[Lin and Donnelly(1966)]{lin1966gas}
T.~J. Lin and H.~G. Donnelly.
\newblock Gas bubble entrainment by plunging laminar liquid jets.
\newblock \emph{AIChE Journal}, 12\penalty0 (3):\penalty0 563--571, 1966.

\bibitem[Liu et~al.(2022)Liu, Gao, and Hu]{liu2022consistent}
C.~Liu, R.~Gao, and C.~Hu.
\newblock A consistent mass--momentum flux computation method for the
  simulation of plunging jet.
\newblock \emph{Physics of Fluids}, 34\penalty0 (3):\penalty0 032114, 2022.

\bibitem[Liu et~al.(2023)Liu, Zeng, Ren, Li, Zhang, Yang, Li, Yang, Su, Zhu,
  et~al.]{liu2023grounding}
S.~Liu, Z.~Zeng, T.~Ren, F.~Li, H.~Zhang, J.~Yang, C.~Li, J.~Yang, H.~Su,
  J.~Zhu, et~al.
\newblock Grounding dino: Marrying dino with grounded pre-training for open-set
  object detection.
\newblock \emph{arXiv preprint arXiv:2303.05499}, 2023.

\bibitem[Lorenceau et~al.(2004)Lorenceau, Qu{\'e}r{\'e}, and
  Eggers]{lorenceau2004air}
{\'E}lise Lorenceau, David Qu{\'e}r{\'e}, and Jens Eggers.
\newblock Air entrainment by a viscous jet plunging into a bath.
\newblock \emph{Physical review letters}, 93\penalty0 (25):\penalty0 254501,
  2004.

\bibitem[L’vov et~al.(2008)L’vov, Pomyalov, Procaccia, and
  Govindarajan]{l2008random}
Victor~S L’vov, Anna Pomyalov, Itamar Procaccia, and Rama Govindarajan.
\newblock Random vortex-street model for a self-similar plane turbulent jet.
\newblock \emph{Physical review letters}, 101\penalty0 (9):\penalty0 094503,
  2008.

\bibitem[Ma et~al.(2010)Ma, Oberai, Drew, Lahey~Jr, and
  Moraga]{ma2010quantitative}
J~Ma, AA~Oberai, DA~Drew, RT~Lahey~Jr, and FJ~Moraga.
\newblock A quantitative sub-grid air entrainment model for bubbly
  flows--plunging jets.
\newblock \emph{Computers \& Fluids}, 39\penalty0 (1):\penalty0 77--86, 2010.

\bibitem[Maxworthy et~al.(1996)Maxworthy, Gnann, K{\"u}rten, and
  Durst]{maxworthy1996experiments}
T~Maxworthy, C~Gnann, M~K{\"u}rten, and F~Durst.
\newblock Experiments on the rise of air bubbles in clean viscous liquids.
\newblock \emph{Journal of fluid mechanics}, 321:\penalty0 421--441, 1996.

\bibitem[McKeogh and Ervine(1981)]{mckeogh1981air}
E.J. McKeogh and D.A. Ervine.
\newblock Air entrainment rate and diffusion pattern of plunging liquid jets.
\newblock \emph{Chemical Engineering Science}, 36\penalty0 (7):\penalty0
  1161--1172, 1981.

\bibitem[Miwa et~al.(2018)Miwa, Moribe, Tsutsumi, and
  Hibiki]{miwa2018experimental}
S.~Miwa, T.~Moribe, K.~Tsutsumi, and T.~Hibiki.
\newblock Experimental investigation of air entrainment by vertical plunging
  liquid jet.
\newblock \emph{Chemical Engineering Science}, 181:\penalty0 251--263, 2018.

\bibitem[Qu et~al.(2013)Qu, Goharzadeh, Khezzar, and Molki]{qu2013experimental}
X.~Qu, A.~Goharzadeh, L.~Khezzar, and A.~Molki.
\newblock Experimental characterization of air-entrainment in a plunging jet.
\newblock \emph{Experimental Thermal and Fluid Science}, 44:\penalty0 51--61,
  2013.

\bibitem[Roy et~al.(2013)Roy, Maiti, and Das]{roy2013visualisation}
A.~K. Roy, B.~Maiti, and P.~K. Das.
\newblock Visualisation of air entrainment by a plunging jet.
\newblock \emph{Procedia Engineering}, 56:\penalty0 468--473, 2013.

\bibitem[Schindelin et~al.(2012)Schindelin, Arganda-Carreras, Frise, Kaynig,
  Longair, Pietzsch, Preibisch, Rueden, Saalfeld, Schmid,
  et~al.]{Schindelin_NatureMethods2012fiji}
Johannes Schindelin, Ignacio Arganda-Carreras, Erwin Frise, Verena Kaynig, Mark
  Longair, Tobias Pietzsch, Stephan Preibisch, Curtis Rueden, Stephan Saalfeld,
  Benjamin Schmid, et~al.
\newblock Fiji: an open-source platform for biological-image analysis.
\newblock \emph{Nature methods}, 9\penalty0 (7):\penalty0 676--682, 2012.

\bibitem[Sene(1988)]{sene1988air}
K.J. Sene.
\newblock Air entrainment by plunging jets.
\newblock \emph{Chemical Engineering Science}, 43\penalty0 (10):\penalty0
  2615--2623, 1988.

\bibitem[Speirs et~al.(2018)Speirs, Pan, Belden, and Truscott]{speirs2018water}
Nathan~B Speirs, Zhao Pan, Jesse Belden, and Tadd~T Truscott.
\newblock The water entry of multi-droplet streams and jets.
\newblock \emph{Journal of Fluid Mechanics}, 844:\penalty0 1084--1111, 2018.

\bibitem[Suciu and Smigelschi(1976)]{suciu1976gas}
G.D. Suciu and O.~Smigelschi.
\newblock Gas absorption by turbulent liquid plunging jets.
\newblock \emph{Chemical Engineering Sciences -- Shorter Communications},
  32:\penalty0 889--896, 1976.

\bibitem[Tsai(1985)]{Tsai_1985CompVisionthresholding}
Wen-Hsiang Tsai.
\newblock Moment-preserving thresolding: A new approach.
\newblock \emph{Computer Vision, Graphics, and Image Processing}, 29\penalty0
  (3):\penalty0 377--393, 1985.
\newblock ISSN 0734-189X.
\newblock \doi{https://doi.org/10.1016/0734-189X(85)90133-1}.

\bibitem[van~de Donk(1981)]{van1981water}
J.~A.~C. van~de Donk.
\newblock \emph{Water aeration with plunging jets}.
\newblock PhD thesis, Delf University of Technology Delf, The Netherland, 1981.

\bibitem[Van~de Sande and Smith(1973)]{van1973surface}
E~Van~de Sande and John~M Smith.
\newblock Surface entrainment of air by high velocity water jets.
\newblock \emph{Chemical Engineering Science}, 28\penalty0 (5):\penalty0
  1161--1168, 1973.

\bibitem[Van~de Sande and Smith(1976)]{van1976jet}
E~Van~de Sande and John~M Smith.
\newblock Jet break-up and air entrainment by low velocity turbulent water
  jets.
\newblock \emph{Chemical Engineering Science}, 31\penalty0 (3):\penalty0
  219--224, 1976.

\bibitem[Wang et~al.(2018)Wang, Slamet, Zhang, and Chanson]{wang2018intrusive}
H.~Wang, N.~S. Slamet, G.~Zhang, and H.~Chanson.
\newblock Intrusive measurements of air-water flow properties in highly
  turbulent supported plunging jets and effects of inflow jet conditions.
\newblock \emph{Chemical Engineering Science}, 177:\penalty0 245--260, 2018.

\bibitem[Zhu et~al.(2000)Zhu, O{\u{g}}uz, and Prosperetti]{zhu2000mechanism}
Y.~Zhu, H.~N. O{\u{g}}uz, and A.~Prosperetti.
\newblock On the mechanism of air entrainment by liquid jets at a free surface.
\newblock \emph{Journal of Fluid Mechanics}, 404:\penalty0 151--177, 2000.
\end{thebibliography}

\end{document}